\documentclass[12pt]{article}
\usepackage{graphicx}

\def\mystrut{\vrule height 3.5ex depth 2.2ex width 0pt}
\def \beq{\begin{equation}}
\def \eeq{\end{equation}}
\def\bea{\begin{eqnarray}}
\def\eea{\end{eqnarray}}

\def\red{
}
\def\black{
}

\def\URLtilde{\lower0.2em\hbox{$\tilde{\phantom{a}}$}}
\def\mycomm#1{\hfill\break\strut\kern-3em{\red\tt ====> #1\black}\hfill\break}
\def\mycommNL#1{\strut\kern-3em{\red\tt ====> #1\black}\hfill\break}

\begin{document}
\thispagestyle{empty}
\rightline{EFI 07-16}
\rightline{TAUP 2857/07}
\rightline{WIS/09/07-JUNE-DPP}
\rightline{ANL-HEP-PR-07-41}
\vskip3cm

\centerline{\large \bf Predictions for masses of $\Xi_b$ baryons}
\bigskip

\centerline{Marek Karliner$^a$, Boaz Keren-Zur$^a$, Harry J. Lipkin$^{a,b,c}$,
and Jonathan L. Rosner$^d$}
\medskip

\centerline{$^a$ {\it School of Physics and Astronomy}}
\centerline{\it Raymond and Beverly Sackler Faculty of Exact Sciences}
\centerline{\it Tel Aviv University, Tel Aviv 69978, Israel}
\medskip

\centerline{$^b$ {\it Department of Particle Physics}}
\centerline{\it Weizmann Institute of Science, Rehovoth 76100, Israel}
\medskip

\centerline{$^c$ {\it High Energy Physics Division, Argonne National
Laboratory}}
\centerline{\it Argonne, IL 60439-4815, USA}
\medskip

\centerline{$^d$ {\it Enrico Fermi Institute and Department of Physics}}
\centerline{\it University of Chicago, 5640 S. Ellis Avenue, Chicago, IL
60637, USA}
\bigskip
\strut
\bigskip
\strut
\bigskip

\centerline{\bf ABSTRACT}
\bigskip

\begin{quote}
The recent observation by CDF of $\Sigma_b^{\pm}$ ($uub$ and $ddb$) 
baryons within 2~MeV of the predicted $\Sigma_b - \Lambda_b$ splitting
has provided strong confirmation for the theoretical approach based
on modeling the color hyperfine interaction. We now apply this approach to 
predict the masses of the $\Xi_b$ family of baryons with quark content $usb$
and $dsb$ -- the ground state $\Xi_b$ at 5790 to 5800 MeV, and the excited
states $\Xi_b^\prime$ and $\Xi_b^*$.  The main source of uncertainty is the
method used to estimate the mass difference $m_b - m_c$ from known hadrons.
We verify that corrections due to the details of the interquark potential
and to $\Xi_b$--$\Xi_b^\prime$ mixing are small. 
\end{quote}

\vfill
\leftline{PACS codes: 14.20.Mr, 12.40.Yx, 12.39.Jh, 11.30.Hw}

\vfill\eject

\begin{section}{Introduction}

For many years the only confirmed baryon with a $b$ quark was the isospin-zero
$\Lambda_b$.  A recent measurement of its mass by the CDF Collaboration is
$M(\Lambda_b) = 5619.7 \pm 1.2 \pm 1.2$ MeV \cite{Acosta:2005mq}.  It has the
quark content $\Lambda_b = bud$, where the $ud$ pair has spin and isospin
$S(ud) = I(ud) =0$.  Now CDF has reported the observation of candidates for the
$\Sigma_b^\pm$ and $\Sigma_b^{*\pm}$ \cite{CDFsigb} with masses consistent with
quark model predictions 
\cite{earlier,Jenkins:1996de, Karliner:2003sy,Ebert:2005xj,Rosner:2006yk},

\bea
M(\Sigma_b^-) - M(\Lambda_b) = 195.5^{+1.0}_{{-}1.0}\,({\rm stat.}) \pm
0.1\, \hbox{(syst.) MeV}
\nonumber\\
\\
M(\Sigma_b^+) - M(\Lambda_b) = 188.0^{+2.0}_{{-}2.3}\,({\rm stat.}) \pm
0.1\, \hbox{(syst.) MeV}
\nonumber
\eea
with isospin-averaged mass difference
$M(\Sigma_b) - M(\Lambda_b) = 192$ MeV,
to be compared with the prediction \cite{Karliner:2003sy,Karliner:2006sy}
$M_{\Sigma_b} - M_{\Lambda_b} = 194 \,{\rm MeV}$.

The $\Sigma_b^\pm$ states consist of a light
quark pair $uu$ or $dd$ with $S=I=1$ coupled with the $b$ quark to $J=1/2$,
while in the $\Sigma_b^{*\pm}$ states the light quark pair and the $b$ quark
are coupled to $J=3/2$.  The CDF sensitivity appears adequate to detect
further heavy baryons, such as those with quark content $bsu$ or $bsd$.  The
S-wave levels of these quarks
consist of the $J=1/2$ states $\Xi_b^{0,-}$ and ${\Xi'}_b^{(0,-)}$ and the
$J=3/2$ states $\Xi_b^{*(0,-)}$.  In this paper we predict the masses of these
states and estimate the dependence of the predictions on the
form of the interquark potential.  This exercise has been applied previously to
hyperfine splittings of known heavy hadrons \cite{Keren-Zur:2007vp}.

We discuss the predictions for $M(\Xi_b)$ in Section 2, starting with an
extrapolation from $M(\Xi_c)$ without correction for hyperfine (HF)
interaction and then estimating this correction.  In the $\Xi_b$ the light
quarks are approximately in a state with $S=0$, while another heavier state
${\Xi'}_b$ is expected in which the light quarks mainly have $S=1$. There is
also a state $\Xi_b^*$ expected with light-quark spin 1 and total $J=3/2$.
Predictions for ${\Xi'}_b$ and $\Xi_b^*$ masses are discussed in Section 3.
We estimate the effect of mixing between light-quark spins $S=0$ and 1 in
Section 4, while Section 5 summarizes.
\end{section}

\begin{section}{$\Xi_b$ mass prediction}
In our model
the mass of a hadron is given by the sum 
of the constituent quark masses plus the color-hyperfine (HF) interactions:
\begin{equation}
V^{HF}_{ij}=v\frac{\vec{\sigma_i}\cdot\vec{\sigma_j}}{m_im_j}\langle\delta(r_{ij})\rangle
\end{equation}
where the $m_i$ is the mass of the $i$'th constituent quark, $\sigma_i$ its spin, $r_{ij}$ the distance between the quarks and $v$ is the interaction strength. We shall neglect the mass differences between $u$ and $d$
constituent quarks, writing $u$ to stand for either $u$ or $d$. All the hadron masses (the ones used and the predictions) are for isospin-averaged baryons and are given in MeV.

The $s$ and $u$ quarks in $\Xi_q$ ($q$ standing for $c$ or $b$) are assumed to be in relative spin 0 and the total mass is given by the expression:
\begin{eqnarray}
\Xi_q=m_q+m_s+m_u-\frac{3v\langle \delta(r_{us}) \rangle}{m_um_s}
\end{eqnarray}
The $\Xi_b$ mass can thus be predicted using the known $\Xi_c$ baryon mass as a
starting point and adding the corrections due to mass differences and HF interactions:
\begin{eqnarray}
\Xi_b&=&\Xi_c + (m_b - m_c) -\frac{3v}{m_um_s}\Bigg( \langle \delta(r_{us})
\rangle_{\Xi_b} -   \langle \delta(r_{us}) \rangle_{\Xi_c} \Bigg)
\end{eqnarray}

The experimentally determined masses for the charmed-strange baryons $\Xi_c$, $\Xi'_c$, and
$\Xi^*_c$ are \cite{PDG2006}:
\begin{equation}
\Xi_c=2469.5\pm0.5~\rm{MeV} \qquad
\Xi_c'=2577\pm4~\rm{MeV} \qquad
\Xi_c^*=2646.3\pm1.8~\rm{MeV} ~.
\end{equation}
\begin{subsection}{Constituent quark mass difference}
The mass difference $(m_b - m_c)$ can be obtained from experimental data using one of the following
expressions:

\begin{itemize}

\item
We can simply take the difference of the masses of the $\Lambda_q$ baryons,
ignoring the differences in the HF interaction:
\begin{eqnarray}
m_b - m_c = \Lambda_b - \Lambda_c = 3333.2 \pm 1.2~.
\label{eq_lambda_b_lambda_c}
\end{eqnarray}

\item 
We can use the spin averaged masses of the $\Lambda_q$ and $\Sigma_q$ baryons:
\begin{eqnarray}
m_b - m_c = \Bigg(\frac{2 \Sigma_b^* + \Sigma_b+ \Lambda_b}{4}
- \frac{2\Sigma_c^* + \Sigma_c + \Lambda_c}{4}\Bigg) = 3330.4 \pm 1.8~.
\label{eq_sigma_b_sigma_c}
\end{eqnarray}

\item 
Since the $\Xi_q$ baryon has strangeness 1, it might be better to use masses of
mesons with $S=1$:
\begin{eqnarray}
m_b - m_c = \Bigg(\frac{3B_s^* + B_s}{4} - \frac{3D_s^* + D_s}{4}\Bigg)
 = 3324.6 \pm 1.4~.
\label{eq_B_s_D_s}
\end{eqnarray}

\end{itemize}
\end{subsection}

\begin{subsection}{HF interaction correction}

The HF interaction correction can also be based on $\Xi_c$ baryon experimental
data:
\begin{eqnarray}
\frac{v}{m_um_s}\Bigg( \langle \delta(r_{us}) \rangle_{\Xi_b} -   \langle \delta(r_{us}) \rangle_{\Xi_c} \Bigg) 
&=&\frac{v\langle \delta(r_{us}) \rangle_{\Xi_c}}{m_um_s}\Bigg(\frac{\langle \delta(r_{us}) \rangle_{\Xi_b}}{\langle \delta(r_{us}) \rangle_{\Xi_c}}-1 \Bigg) \\ \nonumber
&=&\frac{2\Xi_c^*+\Xi_c'-3\Xi_c}{12}\Bigg(\frac{\langle \delta(r_{us}) \rangle_{\Xi_b}}{\langle \delta(r_{us}) \rangle_{\Xi_c}}-1 \Bigg)\\ \nonumber
&=&\Bigg(\frac{\langle \delta(r_{us}) \rangle_{\Xi_b}}{\langle \delta(r_{us}) \rangle_{\Xi_c}}-1 \Bigg)(38.4\pm0.5)~\rm{MeV}
\label{eq_HF_correction}
\end{eqnarray}
However, this expression requires the calculation of the $\delta$ function expectation values. These were calculated  using 3-body wavefunctions obtained by a variational method as described in \cite{Keren-Zur:2007vp}. The only input required for these calculations is the shape of confining potential, because the coupling constants cancel out when taking the ratio of the $\delta$ function expectation values. The potentials considered in this work are the linear, Coulomb and Cornell (Coulomb + linear) potentials. We also wrote down the results obtained without the HF corrections. Note that in the case of the Cornell potential we have an additional parameter, which determines the ratio between the strengths of the linear and Coulombic parts of the potential. In these calculations we used the parameters extracted in \cite{Cornell:1980} from analysis of quarkonium spectra (or $K=0.45$ when using the parameterization described in \cite{Keren-Zur:2007vp}).

As a test case we compared the values obtained from experimental data and variational calculations for the ratio of contact probabilities in $\Xi$ and $\Xi_c$. 

\begin{eqnarray}
\frac{2\Xi_c^*+\Xi_c'-3\Xi_c}{2(\Xi^*-\Xi)}=
\frac{\displaystyle{\frac{6v\langle \delta(r_{us})\rangle_{\Xi_c}}{m_um_s}}}
{\displaystyle{\frac{6v\langle \delta(r_{us})\rangle_{\Xi}}{m_um_s}}}=
\frac{\langle \delta(r_{us})\rangle_{\Xi_c}}{\langle \delta(r_{us})\rangle_{\Xi}}
\label{eq_Xi_Xic}
\end{eqnarray}
\end{subsection}
The results given in Table \ref{tab_Xi_Xic} show good agreement between data
and theoretical predictions using the Cornell potential.
\begin{table}[!htbp]
	\centering
		\begin{tabular}{ccc} \hline \hline
\mystrut		&${\langle \delta(r_{us})\rangle_{\Xi_c}}/{\langle
 \delta(r_{us})\rangle_{\Xi}}$ \\ \hline 
\mystrut Experimental data~\cite{PDG2006} & $1.071\pm0.069$ \\ \hline
\mystrut Linear                   & $1.022\pm0.072$ \\
\mystrut Coulomb                  & $1.487\pm0.002$ \\ 
\mystrut Cornell                  & $1.063\pm0.047$ \\ \hline \hline
		\end{tabular}
\caption{\small{Comparison between experimental data and predictions of the
ratio of $u$ and $s$ contact probabilities in $\Xi$ and $\Xi_c$
(Eq.\ (\ref{eq_Xi_Xic})).}}		
\label{tab_Xi_Xic}		
\end{table}

The final predictions for the $\Xi_b$ mass with the different assumptions
regarding the constituent quark mass differences and the confinement potentials
are given in Table \ref{tab_Xib}.  From previous experience we know that the
predictions of the Coulomb potential model show a very strong dependence on
the quark masses which is not observed in the data, hence one should probably
give these predictions less weight. Ignoring the Coulomb potential, one gets a
prediction for the $\Xi_b$ mass in the range of 5790 - 5800 MeV.
\begin{table}[!htbp]
	\centering
\begin{tabular}{cccc} \hline \hline
\mystrut $m_b-m_c =$ & $\Lambda_b-\Lambda_c$ 
                     & ${\Sigma_b}-{\Sigma_c}$ 
                     & ${B_s}-{D_s}$          \\  
& Eq.~(\ref{eq_lambda_b_lambda_c})
& Eq.~(\ref{eq_sigma_b_sigma_c})&eq.~(\ref{eq_B_s_D_s}) \\ \hline 
\mystrut No HF   correction & $5803\pm2$   & $5800\pm2$  & $5794\pm2$ \\
\mystrut Linear             & $5801\pm11$  & $5798\pm11$ & $5792\pm11$ \\
\mystrut Coulomb            & $5778\pm2$   & $5776\pm2$  & $5770\pm2$ \\
\mystrut Cornell            & $5799\pm7$   & $5796\pm7$  & $5790\pm7$ \\
\hline \hline
\end{tabular}
\caption{\small{Predictions for the $\Xi_b$ mass with various confining
potentials and methods of obtaining the quark mass difference $m_b-m_c$ }}		
\label{tab_Xib}		
\end{table}

\end{section}

\begin{section}{$\Xi_b^*$, $\Xi_b'$ mass prediction}
\begin{subsection}{Spin averaged mass (2$\Xi_b^*+\Xi_b'$)/3}
The $s$ and $u$ quarks of the $\Xi_q^*$ and $\Xi_q'$ baryons are assumed to be
in a state of relative spin 1.  We then find
\begin{eqnarray}
\Xi_q^*&=&m_q+m_s+m_u+v\Bigg( \frac{\langle \delta(r_{qs}) \rangle}{m_qm_s}
       +\frac{\langle \delta(r_{qu}) \rangle}{m_qm_u}
       +\frac{\langle \delta(r_{us}) \rangle}{m_um_s}\Bigg)\\ \nonumber
\Xi_q'&=&m_q+m_s+m_u+v\Bigg( \frac{-2\langle \delta(r_{qs}) \rangle}{m_qm_s}
       +\frac{-2\langle \delta(r_{qu}) \rangle}{m_qm_u}
       +\frac{\langle \delta(r_{us}) \rangle}{m_um_s}\Bigg)\\ \nonumber
\end{eqnarray}
The spin-averaged mass of these two states can be expressed as 
\begin{eqnarray}
\frac{2\Xi_q^*+\Xi_q'}{3} = m_q + m_s + m_u
+ \frac{v\langle \delta(r_{us}) \rangle}{m_um_s}~,
\end{eqnarray}
and as for the $\Xi_b$ case, the following prediction can be given:
\begin{eqnarray}
\frac{2\Xi_b^*+\Xi_b'}{3}=\frac{2\Xi_c^*+\Xi_c'}{3} + (m_b - m_c) +
\frac{2\Xi_c^*+\Xi_c'-3\Xi_c}{12} 
\Bigg(\frac{ \langle \delta(r_{us}) \rangle_{\Xi_b}}
{\langle \delta(r_{us}) \rangle_{\Xi_c}}-1\Bigg)~.
\end{eqnarray}
The predictions obtained using the same methods described above are given in
Table \ref{tab_Xib_star}. 
In this case it is clear that the effect of the HF correction is negligible.
Thus the difference between the spin averaged mass $(2\Xi_b^*+\Xi_b')/3$ and
$\Xi_b$ is roughly $150-160$ MeV.

\begin{table}[!htbp]
	\centering
		\begin{tabular}{cccc} \hline \hline
\mystrut $m_b-m_c =$ & $\Lambda_b-\Lambda_c$ 
                     & ${\Sigma_b}-{\Sigma_c}$ 
                     & ${B_s}-{D_s}$          \\  
&Eq.~(\ref{eq_lambda_b_lambda_c})&Eq.~(\ref{eq_sigma_b_sigma_c})
&Eq.~(\ref{eq_B_s_D_s})\\ \hline 
\mystrut No HF correction     & $5956\pm3$ & $5954\pm3$ & $5948\pm3$ \\
\mystrut Linear               & $5957\pm4$ & $5954\pm4$ & $5948\pm4$ \\
\mystrut Coulomb              & $5965\pm3$ & $5962\pm3$  & $5956\pm3$ \\
\mystrut Cornell              & $5958\pm3$ & $5955\pm3$ & $5949\pm3$ \\
\hline \hline
		\end{tabular}
		\caption{\small{Predictions for the spin averaged $\Xi_b'$ and
$\Xi_b^*$ masses with various confining potentials and methods of obtaining the
quark mass difference $m_b-m_c$}}		
\label{tab_Xib_star}
\end{table}
\end{subsection}

\begin{subsection}{$\Xi_b^*-\Xi_b'$}
This mass difference is more difficult to predict, but it will be small due to
the large mass of the $b$ quark.
\begin{equation}
\Xi_q^*-\Xi_q'=3v\Bigg(\frac{\langle \delta(r_{qs}) \rangle}{m_qm_s}
+\frac{\langle \delta(r_{qu}) \rangle}{m_qm_u}\Bigg)
\end{equation}
We can once again use the $\Xi_c$ hadron masses: 
\begin{eqnarray}
\frac{\Xi_b^*-\Xi_b'}{\Xi_c^*-\Xi_c'}=
\frac
{\displaystyle{3v\Bigg(
\frac{\langle \delta(r_{bs}) \rangle}{m_bm_s}
+\frac{\langle \delta(r_{bu}) \rangle}{m_bm_u}\Bigg)}}
{\displaystyle{3v\Bigg(
\frac{\langle \delta(r_{cs}) \rangle}{m_cm_s}
+\frac{\langle \delta(r_{cu}) \rangle}{m_cm_u}\Bigg)}}=
\frac{m_c}{m_b}
\frac{\displaystyle{\Bigg ( \langle \delta(r_{bs}) \rangle_{\Xi_b}
 + \frac{m_s}{m_u} \langle \delta(r_{bu}) \rangle_{\Xi_b} \Bigg)}}
     {\displaystyle{\Bigg ( \langle \delta(r_{cs}) \rangle_{\Xi_c}
 + \frac{m_s}{m_u} \langle \delta(r_{cu}) \rangle_{\Xi_c} \Bigg)}}
\end{eqnarray}
 
This expression is strongly dependent on the confinement model. In the results
given in Table \ref{tab_Xibstar_Xibprime} we have used
$\displaystyle{\frac{m_s}{m_u}=1.5\pm0.1}$,
$\displaystyle{\frac{m_b}{m_c}=2.95\pm0.2}$.

\begin{table}[!htbp]
	\centering
\begin{tabular}{cc} \hline \hline
\mystrut                     & $\Xi_b^*-\Xi_b'$      \\ \hline 
\mystrut No HF correction      & $24\pm2$  \\
\mystrut Linear                & $28\pm6$ \\
\mystrut Coulomb               & $36\pm7$  \\
\mystrut Cornell               & $29\pm6$  \\ \hline \hline
\end{tabular}
\caption{\small{Predictions for the mass difference between $\Xi_b^*$ and $\Xi_b'$ with various confining potentials.}}
\label{tab_Xibstar_Xibprime}
\end{table}

\end{subsection}
\end{section}

\begin{section}{Effect of light-quark spin mixing on $\Xi_b$ and $\Xi'_b$}

In estimates up to this point we have assumed that the light-quark spins
in $\Xi_b$ and $\Xi'_b$ are purely $S=0$ and $S=1$, respectively.  The
differing hyperfine interactions between the $b$ quark and nonstrange or
strange quarks leads to a small admixture of the opposite-$S$ state in each
mass eigenstate \cite{Maltman:1980er,Lipkin:1981,Rosner:1981yh,Rosner:1992qa}.
The effective hyperfine Hamiltonian may be written \cite{Rosner:1981yh,%
Rosner:1992qa}
\hfill\break
\begin{eqnarray}
H_{\rm eff} &=& M_0 + \lambda(\sigma_u \cdot \sigma_s + \alpha \sigma_u \cdot
\sigma_b + \beta \sigma_s \cdot \sigma_b)~,
\end{eqnarray}
where $M_0$ is the sum of spin independent terms, $\lambda \sim 1/(m_u m_s)$, $\alpha = m_s/m_b$, and $\beta = m_u/m_b$.
The calculation of $M_{3/2}$ is straightforward, as the expectation value of
each $\sigma_i \cdot \sigma_j$ in the $J=3/2$ state is 1.
For the $J=1/2$ states one has to diagonalize the $2 \times 2$ matrix
\beq
{\cal M}_{1/2} = \left[ \begin{array}{c c}
 M_0 - 3 \lambda & \lambda \sqrt{3} (\beta - \alpha) \cr
\lambda \sqrt{3} (\beta - \alpha) & M_0 + \lambda (1 - 2 \alpha - 2 \beta)
\end{array} \right]~.
\eeq

The eigenvalues of $H_{\rm eff}$ are thus
\begin{eqnarray}
M_{3/2} & = & M_0 + \lambda(1 + \alpha + \beta)~,\\
M_{1/2,\pm} & = & M_0 + \lambda[ -(1 + \alpha + \beta) \nonumber \\
  & \pm & 2\lambda(1 + \alpha^2 + \beta^2 - \alpha - \beta - \alpha \beta)^{1/2}~.
\end{eqnarray}

In the absence of mixing $(\alpha = \beta)$ one would have $M_{3/2} =
M_0 + \lambda(1 + 2 \alpha)$, $M_{1/2,+} = M_0 + \lambda (1 - 4 \alpha)$,
and $M_{1/2,-} = M_0 - 3 \lambda$.  

To see the effect of mixing, we rewrite the expression for 
$M_{1/2,\pm}$,
\beq
M_{1/2,\pm} = M_0 - \lambda(1 + \alpha + \beta) \pm
 2\lambda\left[\left(1 - {\alpha + \beta\over 2}\right)^2
 + \frac{3}{4}(\alpha - \beta)^2 \right]^{1/2}
\eeq
The effect of the mixing is seen in the term $\frac{3}{4}(\alpha -
\beta)^2$. Expanding $M_{1/2,\pm}$ to second order in small $\alpha-\beta$, 
we obtain
\beq
M_{1/2,\pm} \approx \left(\hbox{terms without mixing}\right) 
\pm \lambda\cdot {\displaystyle {3\over4\strut}
(\alpha-\beta)^2\over \displaystyle 1 -  {\alpha + \beta\strut\over 2}}
\eeq

For $m_u = 363$ MeV, $m_s=538$ MeV, 
and $m_b = 4900$ MeV \cite{Gasiorowicz:1981jz},
one has $\alpha \simeq 0.11$, $\beta \simeq 0.07$, while the discussion in the
previous section implies $\lambda \simeq 40$ MeV [Eq.\
(\ref{eq_HF_correction})]. 
Hence the effect of mixing on our predictions is negligible, amounting to 
$\pm 0.04$ MeV.

Since we use the $\Xi_c$ and $\Xi_c^\prime$ masses as input for $\Xi_b$,
it is also important to check the mixing effects on the former. Since
$m_b/m_c \sim 3$, this amounts to changing in the expressions above
$\alpha \rightarrow 3 \alpha$,
$\beta \rightarrow 3 \beta$. The corresponding effect of mixing
on $\Xi_c$ and $\Xi_c^\prime$ is $\sim 0.5$ MeV, still
negligible.

\end{section}

\begin{section}{Summary}

We have shown that predictions for $M(\Xi_b)$ based on the masses of $\Xi_c$,
$\Xi'_c$, and $\Xi^*_c$ lie in the range of 5790 to 5800 MeV, depending on how
the mass difference $m_b - m_c$ is estimated.  Wave function differences tend
to affect these predictions by only a few MeV.  The spin-averaged mass of
the states $\Xi'_b$ and $\Xi^*_b$ is predicted to lie  around 150 to 160 MeV
above $M(\Xi_b)$, while the hyperfine splitting between $\Xi'_b$ and $\Xi^*_b$
is predicted to lie in the rough range of 20 to 30 MeV.  We look forward to
the verification of these predictions in experiments at the Fermilab Tevatron
and the CERN Large Hadron Collider.  

Note added:  After this work was completed we received notice of the $\Xi_b^-$
observation at the Fermilab Tevatron in the $J/\psi \Xi^-$ decay mode by the D0
Collaboration \cite{Abazov:2007ub}. After the first version of this paper
appeared \cite{v1}, the CDF Collaboration released their $\Xi_b^-$  results in
the same decay channel \cite{Litvintsev:2007}. The reported masses, Gaussian
widths (due to instrumental resolution), and significances of the signal are
summarized in Table~\ref{tab:xib_exp} and in Fig.~1. 
 The CDF Collaboration also observes a
significant $\Xi_b^- \to \Xi_c^0 \pi^-$ signal with mass consistent with that
found in the $J/\psi \Xi^-$ mode. The D0 mass is consistent with all our
predictions for the isospin-averaged mass, while that of CDF allows us to rule
out the (previously disfavored \cite{Keren-Zur:2007vp}) prediction based on the
Coulomb potential.  Both experiments also agree with a prediction in Ref.\
\cite{Jenkins:1996de}, $M(\Xi_b) = M(\Lambda_b) + (182.7 \pm 5.0)$ MeV $ =
(5802.4 \pm 5.3)$ MeV, where differences in wave function effects were not
discussed and $m_b-m_c$ was taken from baryons only, whereas in our work the
optional value of  $m_b-m_c$ was obtained from $B_s$ and $D_s$ mesons which
contain both heavy and strange quarks, as do $\Xi_b$ and $\Xi_c$. See also
Ref.~\cite{Ebert:2005xj} and Table III therein for a compilation of earlier
predictions for the $\Xi_b$ mass.  That the
value of $m_b-m_c$ obtained from $B$ and $D$ mesons depends  upon the flavor of
the spectator quark was noted in Ref.~\cite{Karliner:2003sy} where Table I shows
that the value is the same for mesons and baryons not containing strange quarks
but different when obtained from $B_s$ and $D_s$ mesons. Some reasons for this
difference were noted and the issue requires further investigation. Here we
have updated the prediction of Ref.~\cite{Jenkins:1996de}  using the recent CDF
\cite{Acosta:2005mq} value of $M(\Lambda_b)$.
\hfill\break
\strut\vskip-1cm

\begin{table}
\caption{Observations of $\Xi_b^- \to J/\psi \Xi^-$ at the
Fermilab Tevatron.  Errors on mass are (statistical, systematic).
\label{tab:xib_exp}}
\begin{center}
\begin{tabular}{c c c} \hline \hline
  & D0 \cite{Abazov:2007ub} & CDF \cite{Litvintsev:2007} \\ \hline
Mass (MeV) & $5774 \pm 11 \pm 15$
 & $5793 \pm 2.4 \pm 1.7$ \\
Width (MeV) & $37 \pm 8$ & $\sim 14$ \\
Significance & $5.5 \sigma$ & $7.8 \sigma$ \\ \hline \hline
\end{tabular}
\end{center}
\hfill\break
\hfill\break
\hfill\break
\hfill\break
\strut\kern-0.7cm
\includegraphics[width=10.0cm,angle=90]{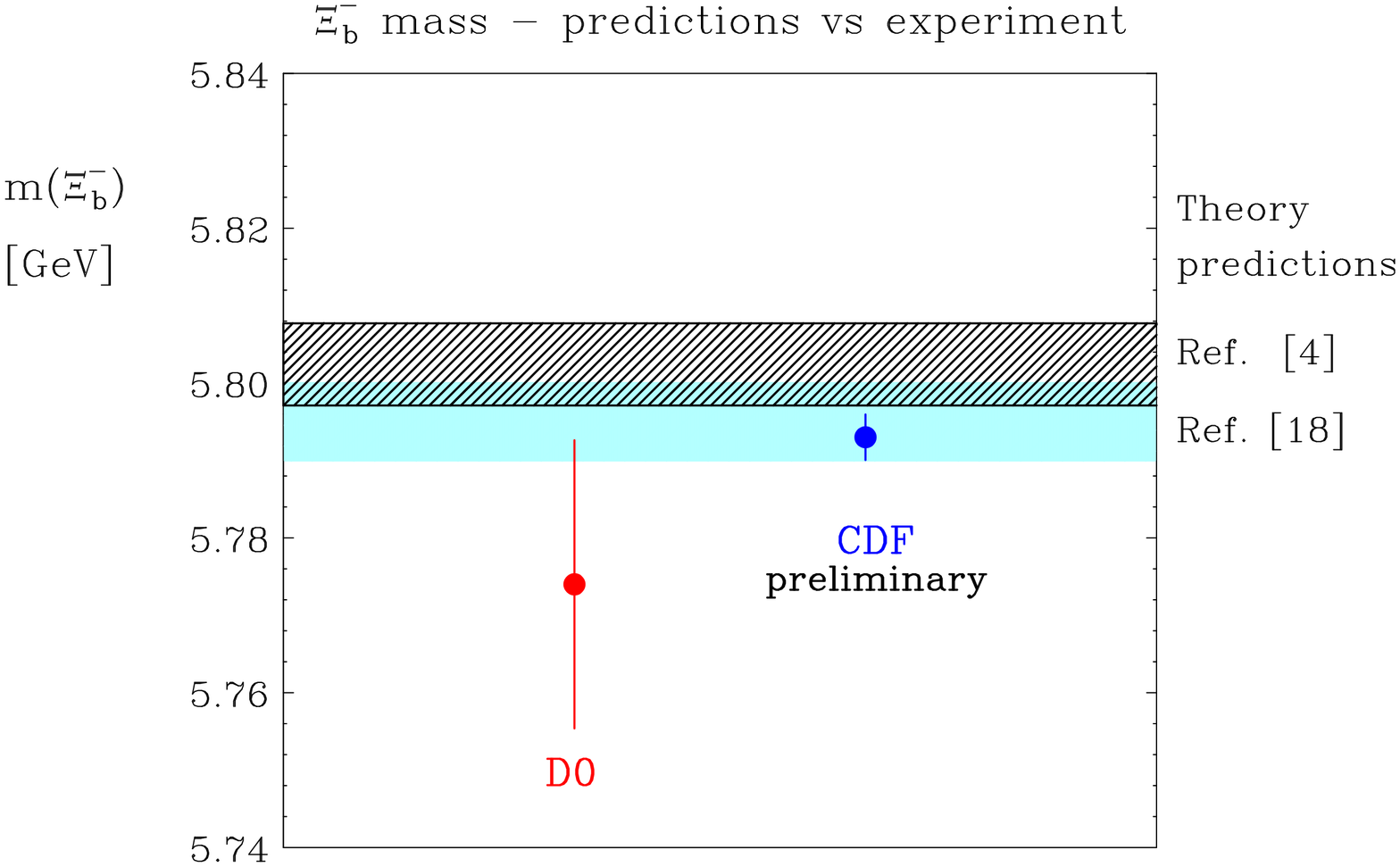}
\hfill\break
\hfill\break
{\small Fig.\ 1 (adapted from \cite{Litvintsev:2007}). 
Comparison of theoretical predictions and experimental 
results for the $\Xi^-_b$ mass from D0 \cite{Abazov:2007ub} and CDF
\cite{Litvintsev:2007}.
The theoretical
predictions are denoted by the two
horizontal bands, corresponding to Refs.~\cite{Jenkins:1996de} and \cite{v1},
respectively.}
\end{table}

\end{section}
\section*{Acknowledgements}
J.L.R. wishes to acknowledge the hospitality of Tel Aviv University during
the early stages of this work.
We thank Dmitry Litvintsev for providing his figure comparing
theoretical predictions with measurements of the $\Xi^-_b$ mass.
This research was supported in part by a grant from Israel Science Foundation
administered by Israel Academy of Science and Humanities. The research of 
H.J.L. was supported in part by the U.S. Department of Energy, Division of
High Energy Physics, Contract DE-AC02-06CH11357.
The work of J.L.R. was supported by the U.S. Department of Energy, Division of
High Energy Physics, Grant No.\ DE-FG02-90ER40560.

\end{document}